\documentclass[11pt]{article}

\usepackage[T1]{fontenc}
\usepackage[utf8]{inputenc}
\usepackage[a4paper,margin=1in]{geometry}
\usepackage{microtype}
\usepackage{amsmath}
\usepackage{booktabs}
\usepackage{array}
\usepackage[hidelinks]{hyperref}
\usepackage{authblk}
\usepackage{url}
\usepackage{cleveref}

\hypersetup{
  pdftitle={Colour Contrast on the Web: A WCAG 2.1 Level AA Compliance Audit of Common Crawl's Top 500 Domains},
  pdfauthor={T E Vaughan},
  pdfsubject={Web accessibility, WCAG, Common Crawl},
  pdfkeywords={WCAG, accessibility, colour contrast, Common Crawl, web archives}
}

\title{Colour Contrast on the Web:\\
A WCAG 2.1 Level AA Compliance Audit of Common Crawl's Top 500 Domains}

\author[1]{Thom Vaughan}
\author[1]{Pedro Ortiz Suarez}
\affil[1]{Common Crawl Foundation}

\date{February 2026}

\begin{document}
\maketitle

\begin{abstract}
We present a large-scale automated audit of WCAG 2.1/2.2 Level AA colour contrast compliance across the 500 most frequently crawled registered domains in Common Crawl's CC-MAIN-2026-08 February 2026 crawl archive. Rather than conducting a live crawl, all page content was sourced from Common Crawl's open WARC archives, ensuring reproducibility and eliminating any load on target web servers.
Our static CSS analysis of 240 homepages identified 4,327 unique foreground/background colour pairings, of which 1,771 (40.9\%) failed to meet the 4.5:1 contrast ratio threshold for normal text. The median per-site pass rate was 62.7\%, with 20.4\% of sites achieving full compliance across all detected colour pairings. These findings suggest that colour contrast remains a widespread accessibility barrier on the most prominent websites, with significant variation across domain categories.
\end{abstract}

\section{Introduction}

Web accessibility is a fundamental aspect of an inclusive internet. The Web Content Accessibility Guidelines (WCAG), maintained by the World Wide Web Consortium (W3C), define success criteria intended to make web content accessible to people with disabilities, including those with low vision, colour vision deficiencies, and other visual impairments \cite{wcag21,wcag22}.

Among the most measurable of these criteria is colour contrast. WCAG 2.1, under Success Criterion 1.4.3 (Contrast, Minimum), requires that the visual presentation of text and images of text have a contrast ratio of at least 4.5:1 for normal text and 3:1 for large text (defined as text at 18 points or larger, or 14 points or larger if bold) \cite{wcag21_understanding_143}.

This criterion is classified as Level AA, the conformance level most commonly targeted by accessibility regulations worldwide, including the European Accessibility Act, Section 508 of the US Rehabilitation Act, and the UK Equality Act.

Previous studies of web accessibility have typically relied on live crawling of websites, which raises concerns about reproducibility (websites change over time), server load (automated audits can strain web infrastructure), and ethical considerations around crawling without explicit consent.

In this study, we take a different approach: we use the open, freely available web archives maintained by Common Crawl \cite{commoncrawl}, a non-profit organisation that performs broad web crawls on a monthly basis and makes the resulting data freely available to the public.
By sourcing all page content from Common Crawl's WARC files, our analysis is fully reproducible. Any researcher can re-run our pipeline against the same CC-MAIN-2026-08 archive \cite{cc_main_2026_08} and obtain identical results. This approach also eliminates any load on the target websites themselves.

This work investigates whether meaningful, deterministic accessibility analysis can be performed directly from web crawl archives without rendering live pages. Using archived HTML from Common Crawl, we evaluate declared foreground and background colour pairings against WCAG contrast thresholds at web scale.

\section{Related Work}

Large-scale measurement of web accessibility has been an active area of research for over a decade. One of the most influential ongoing efforts is the WebAIM Million project \cite{webaim_million}, which annually evaluates the homepages of the top one million websites for WCAG conformance issues using automated tooling. Their reports consistently identify colour contrast as the most common accessibility failure, with over 75\% of homepages exhibiting at least one contrast error in recent analyses. Unlike our approach, WebAIM performs live crawling and browser-based rendering, executing JavaScript and loading external stylesheets.

Academic work has similarly examined accessibility compliance at scale. Vigo et al. \cite{vigo2013quantifying} quantified accessibility barriers across popular websites using automated WCAG testing tools, highlighting persistent structural and visual issues. Kelly et al. \cite{kelly2009accessibility} explored the limitations of automated accessibility evaluation, noting the gap between machine-detectable failures and real user impact. More recently, Alshayban et al. \cite{alshayban2020accessibility} investigated accessibility in modern web frameworks, demonstrating how client-side rendering and JavaScript-heavy architectures introduce new challenges for automated analysis.

From the information retrieval and web measurement perspective, this work builds on a rich literature analysing the web as a large-scale corpus. Studies using Common Crawl and similar archives have examined language distribution, hyperlink structure, structured data extraction, content duplication, and web evolution \cite{meusel2015graph, meusel2014web, commoncrawl}. Web archive–based measurement has also been used to study longitudinal changes in web content and structure without re-crawling live sites. However, accessibility analysis using archived WARC data, rather than live rendering environments, remains relatively underexplored.

Our work differs from prior large-scale accessibility studies in two principal ways. First, it relies exclusively on archived WARC content from a fixed crawl release, ensuring full reproducibility and eliminating load on origin servers. Second, it evaluates declared CSS colour pairings directly from static HTML, rather than analysing computed styles in a rendered DOM. This positions the study at the intersection of accessibility research and reproducible web-scale measurement using open crawl archives.

More broadly, this paper contributes to the growing body of work advocating reproducible web measurement methodologies that operate on shared public datasets rather than bespoke crawls \cite{olston2010web, meyer2008common}. By demonstrating that meaningful accessibility signals can be extracted deterministically from Common Crawl archives, we extend archive-based analysis into the accessibility domain.

\section{Methodology}

\subsection{Domain Selection}
We selected the 500 most frequently crawled registered domains from Common Crawl's CC-MAIN-2026-08 February 2026 crawl archive, as ranked by page captures in Common Crawl's crawl statistics \cite{cc_crawl_statistics}. These statistics are derived from Common Crawl's URL index data and are publicly available.
The domain list spans a diverse cross-section of the web, including blogging platforms (blogspot.com, wordpress.org), reference sites (wikipedia.org, wiktionary.org), technology companies (google.com, microsoft.com, apple.com), educational institutions, government agencies, and e-commerce platforms.

\subsection{Data Retrieval}
For each domain, we queried Common Crawl's Columnar Index to locate an archived capture of the domain's homepage from the CC-MAIN-2026-08 crawl \cite{commoncrawl_columnar_index}. The Columnar Index is a Parquet-based representation of the crawl index stored on Amazon S3, which can be queried via Amazon Athena. A single SQL query across all 500 domains returns the WARC filename, byte offset, and record length for each homepage capture, identifying the exact location of the page content within Common Crawl's distributed WARC archive.

We then fetched the actual HTML content using HTTP byte-range requests to \path{data.commoncrawl.org}, extracting just the relevant WARC record from the larger archive file. This is the standard method for accessing individual pages within Common Crawl's petabyte-scale archive. Each WARC record contains the original HTTP response headers and body, from which we extracted the HTML document.

The index query was configured to select only successful responses (HTTP status 200) with a detected MIME type of \texttt{text/html}, preferring the \texttt{www} subdomain or bare domain over deeper subdomains, and HTTPS over HTTP where available.

\subsection{Colour Extraction}
From each HTML document, we extracted CSS colour declarations using two complementary methods:
\begin{enumerate}
  \item Embedded stylesheets: All \texttt{<style>} block contents were parsed as CSS, extracting rule selectors and their \texttt{color} and \texttt{background-color} declarations. The CSS \texttt{background} shorthand property was also parsed to extract colour components.
  \item Inline styles: All \texttt{style} attributes on HTML elements were parsed for both \texttt{color} and \texttt{background-color} declarations.
\end{enumerate}
This approach captures the static CSS present in the archived HTML. It does not capture styles applied by JavaScript at runtime, styles loaded from external CSS files (which would require additional index lookups and WARC fetches), or styles applied through CSS custom properties (\texttt{var(--name)}). This is a known limitation that we discuss in Section~\ref{sec:limitations}.

Colour values were parsed from all standard CSS colour formats: hexadecimal notation (\texttt{\#RGB}, \texttt{\#RRGGBB}, \texttt{\#RGBA}, \texttt{\#RRGGBBAA}), the \texttt{rgb()} and \texttt{rgba()} functions, the \texttt{hsl()} and \texttt{hsla()} functions, and all 148 CSS named colours. Non-colour values such as \texttt{transparent}, \texttt{inherit}, \texttt{currentColor}, and \texttt{initial} were excluded from analysis.

\subsection{Colour Pairing and Contrast Calculation}
This analysis measures declared colour contrast properties rather than rendered visual presentation.

Extracted colour declarations were paired into foreground/background combinations using the following rules:
\begin{itemize}
  \item When both \texttt{color} and \texttt{background-color} are specified in the same CSS rule or inline style, they form an explicit pairing.
  \item When only \texttt{color} is specified, it is paired with an assumed white background (\texttt{\#FFFFFF}), reflecting the default rendering behaviour of web browsers.
  \item When only \texttt{background-color} is specified, it is paired with assumed black text (\texttt{\#000000}).
\end{itemize}
Identical pairings (same foreground and background RGB values) were deduplicated to avoid counting the same colour combination multiple times.

For each pairing, we calculated the contrast ratio using the formula defined in WCAG 2.1 \cite{wcag21}:
\begin{equation}
\text{Contrast ratio} = \frac{L_{1} + 0.05}{L_{2} + 0.05}
\end{equation}
where $L_{1}$ is the relative luminance of the lighter colour and $L_{2}$ is the relative luminance of the darker colour. Relative luminance is calculated from linearised sRGB values:
\begin{equation}
L = 0.2126 \cdot R_{\text{lin}} + 0.7152 \cdot G_{\text{lin}} + 0.0722 \cdot B_{\text{lin}}
\end{equation}
where each component is linearised from its 8-bit sRGB value using the standard piecewise function.

\subsection{Compliance Assessment}
Each colour pairing was evaluated against two WCAG 2.1 Level AA thresholds:
\begin{itemize}
  \item Normal text (SC 1.4.3): Requires a contrast ratio of at least 4.5:1
  \item Large text (SC 1.4.3): Requires a contrast ratio of at least 3.0:1
\end{itemize}
Since our static analysis does not determine the rendered font size of text elements, we report pass rates against both thresholds. The normal text threshold (4.5:1) is the more stringent criterion and represents the stricter assessment; the large text threshold (3.0:1) shows how results differ when the more lenient standard applies.

\section{Results}

\subsection{Coverage}
Of the 500 domains in our sample, 428 yielded analysable homepage HTML from the CC-MAIN-2026-08 archive. Of these, 240 contained at least one parseable CSS colour declaration. 188 homepages were retrieved successfully but contained no CSS colour data in their embedded or inline styles (these sites likely rely entirely on external stylesheets or JavaScript-injected styles).

\subsection{Overall Compliance}

\begin{table}[ht]
\centering
\begin{tabular}{l r r}
\toprule
Pass rate range & Domains & Percentage \\
\midrule
100\% (fully compliant) & 49 & 20.4\% \\
90--99\% & 1 & 0.4\% \\
75--89\% & 28 & 11.7\% \\
50--74\% & 99 & 41.2\% \\
25--49\% & 33 & 13.8\% \\
0--24\% & 30 & 12.5\% \\
\bottomrule
\end{tabular}
\caption{Distribution of per-site pass rates for normal text (4.5:1 threshold).}
\label{tab:passrate_distribution}
\end{table}

Across 240 domains with colour data, we identified 4,327 unique foreground/background colour pairings. The overall compliance picture:
\begin{itemize}
  \item Mean per-site pass rate (normal text): 59.8\%
  \item Median per-site pass rate (normal text): 62.7\%
  \item Fully compliant sites: 20.4\% (100\% of pairings passing at 4.5:1)
  \item Sites with $>90\%$ pass rate: 20.8\%
  \item Sites with $<50\%$ pass rate: 26.2\%
\end{itemize}

The distribution of per-site pass rates (normal text threshold) is presented in \Cref{tab:passrate_distribution}

\subsection{Analysis by Domain Category}
\begin{table}[ht]
\centering
\begin{tabular}{l r r r r}
\toprule
Category & Domains & Avg pass rate & Median & Compliant \\
\midrule
Research & 1 & 72.2\% & 72.2\% & 0 \\
EU Institutions & 1 & 66.7\% & 66.7\% & 0 \\
E-commerce & 5 & 64.1\% & 66.7\% & 0 \\
Education & 47 & 63.7\% & 64.7\% & 14 \\
News/Media & 10 & 61.7\% & 75.0\% & 1 \\
Other & 134 & 61.5\% & 64.7\% & 30 \\
Open Knowledge & 7 & 52.4\% & 50.0\% & 0 \\
Government & 14 & 49.7\% & 54.5\% & 1 \\
Technology & 10 & 47.7\% & 50.0\% & 1 \\
Hosting/Platform & 11 & 44.6\% & 43.3\% & 2 \\
\bottomrule
\end{tabular}
\caption{Pass rate statistics by domain category, sorted by average pass rate descending.}
\label{tab:category_stats}
\end{table}

We categorised each domain by its primary function (Education, Government, Technology, News/Media, E-commerce, Hosting/Platform, Open Knowledge, Research, and Other). The distribution and pass rate statistics are presented in \Cref{tab:category_stats}.

\subsection{Notable Findings}

\subsubsection{Worst Offenders}
\begin{table}[ht]
\centering
\begin{tabular}{l r r l}
\toprule
Domain & Pass rate & Failing pairings & Worst ratio \\
\midrule
adelaide.edu.au & 0.0\% & 1 & 1.0:1 \\
alberta.ca & 0.0\% & 1 & 1.0:1 \\
af.mil & 0.0\% & 1 & 1.0:1 \\
copernicus.org & 0.0\% & 1 & 4.13:1 \\
github.io & 0.0\% & 1 & 4.02:1 \\
gamer.com.tw & 0.0\% & 1 & 1.0:1 \\
kit.edu & 0.0\% & 1 & 1.0:1 \\
iol.pt & 0.0\% & 1 & 3.77:1 \\
mts.ru & 0.0\% & 1 & 1.0:1 \\
ncl.ac.uk & 0.0\% & 2 & 1.0:1 \\
\bottomrule
\end{tabular}
\caption{Lowest pass-rate domains (normal text threshold).}
\label{tab:worst_offenders}
\end{table}
\Cref{tab:worst_offenders} presents a list of domains that we found had the lowest pass rates for normal text contrast:

\subsubsection{Fully Compliant Sites}
49 domains achieved a 100\% pass rate across all detected colour pairings at the 4.5:1 threshold. \Cref{tab:fully_compliant} presents a list of Selected fully compliant sites

\begin{table}[ht]
\centering
\begin{tabular}{l r l}
\toprule
Domain & Pairings checked & Mean ratio \\
\midrule
desktopnexus.com & 6 & 8.44:1 \\
hatenablog.com & 6 & 13.14:1 \\
baidu.com & 4 & 17.89:1 \\
craigslist.org & 4 & 12.33:1 \\
fu-berlin.de & 4 & 11.21:1 \\
tokyo.lg.jp & 4 & 14.15:1 \\
unt.edu & 4 & 8.54:1 \\
prnewswire.com & 3 & 13.57:1 \\
anu.edu.au & 2 & 19.52:1 \\
google.cn & 2 & 12.77:1 \\
\bottomrule
\end{tabular}
\caption{Selected fully compliant sites with the most pairings checked.}
\label{tab:fully_compliant}
\end{table}

\section{Discussion}

\subsection{The State of Colour Contrast Compliance}
Our findings indicate that colour contrast compliance varies substantially across the web's most prominent domains. With a median per-site pass rate of 62.7\% for normal text, a significant proportion of CSS-declared colour pairings fail to meet the WCAG AA threshold of 4.5:1. However, the interpretation of these numbers requires nuance.

First, not all colour pairings carry equal weight in a user's experience. A site might have several low-contrast pairings defined in CSS that are rarely or never applied to visible text elements. Our static analysis counts all declared pairings equally, without regard to their prominence or frequency of use on the page.

Second, sites that declare fewer colours in embedded CSS tend to show more extreme pass rates (either very high or very low), whilst sites with many colour declarations tend to cluster around the mean. This is partly an artefact of sites with rich embedded stylesheets offering more opportunities for both passing and failing pairings.

\subsection{Category Differences}
The highest-scoring category was Research with a mean pass rate of 72.2\%, whilst Hosting/Platform had the lowest at 44.6\%. These differences likely reflect varying levels of institutional attention to accessibility standards, with sectors subject to regulatory requirements (such as government and education) potentially investing more in compliance.

\subsection{Implications}
These findings highlight that even among the web's most popular domains, colour contrast barriers remain common. For users with low vision or colour vision deficiencies, these barriers can make content difficult or impossible to read. The prevalence of contrast failures across diverse categories of websites underscores the need for continued advocacy, tooling, and potentially regulation to improve web accessibility.

\section{Conclusion}
We conducted a large-scale WCAG 2.1 Level AA colour contrast audit of the 500 most frequently crawled domains in Common Crawl's CC-MAIN-2026-08 archive, analysing 4,327 unique colour pairings across 240 homepages. Our findings reveal that 79.6\% of sites contain at least one colour pairing that fails the 4.5:1 normal text contrast threshold, with a median per-site pass rate of 62.7\%.

Whilst static CSS analysis provides only a partial picture of a page's true accessibility, these results establish a reproducible baseline for tracking colour contrast compliance over time. Future work could extend this analysis to incorporate external stylesheets, JavaScript-rendered styles, and per-element rendering context, building toward a comprehensive picture of colour accessibility on the open web.
Many failing pairings cluster near the WCAG threshold, which suggests that minor colour adjustments could substantially improve compliance without major redesign.

This analysis indicates that meaningful accessibility research can be performed directly on crawl archives, without re-crawling or executing live web content.

Future work includes longitudinal comparison across crawl releases and integration of rendered-style analysis to compare declared and computed accessibility properties.

\section{Limitations}
\label{sec:limitations}
This study has several important limitations:
\begin{enumerate}
  \item Static analysis only: We parsed CSS declarations present in the archived HTML without executing JavaScript. Modern web applications often inject styles dynamically, use CSS-in-JS libraries, or load styles asynchronously. Our analysis therefore captures a lower bound of the total colour declarations on each page.
  \item No external stylesheets: We did not fetch external CSS files (referenced via \texttt{<link>} elements). Many sites define the majority of their styles in external files. Incorporating external stylesheet analysis would require additional index lookups and WARC fetches for each domain, significantly increasing pipeline complexity.
  \item No rendering context: We cannot determine which CSS selectors apply to which rendered elements, what font sizes are in effect, or whether low-contrast elements are actually visible to users. A pairing that fails the contrast test in our analysis might apply only to decorative elements or hidden content.
  \item Homepage bias: We analysed only each domain's homepage. Internal pages may have different colour schemes, templates, or accessibility characteristics.
  \item Snapshot in time: The CC-MAIN-2026-08 crawl captures a single point in time for each page. Websites may have been redesigned before or after the crawl date.
  \item Assumed defaults: When only a foreground colour was specified without a background, we assumed white (\texttt{\#FFFFFF}), and vice versa with black (\texttt{\#000000}). In practice, inherited styles or user agent defaults may produce different pairings.
\end{enumerate}

\section{Reproducibility}
The complete analysis pipeline is available as open-source Python code requiring only Python 3.9+ and no external dependencies beyond the standard library. The pipeline consists of four steps:
\begin{enumerate}
  \item Columnar Index queries via Amazon Athena to locate homepage captures in CC-MAIN-2026-08
  \item WARC byte-range fetches to retrieve archived HTML
  \item CSS colour extraction and WCAG contrast analysis
  \item Aggregate statistics and report generation
\end{enumerate}
All data is sourced from Common Crawl's publicly available archives. Researchers can reproduce this analysis by running the pipeline against the same CC-MAIN-2026-08 crawl data, or adapt it to analyse different crawls or domain sets.

\bibliographystyle{plain}
\bibliography{references}

\end{document}